\documentclass[preprint,12pt, authoryear]{elsarticle}

\usepackage{bm}
\usepackage{amsmath,amssymb,amsfonts}
\usepackage{algorithmic}
\usepackage{graphicx}
\usepackage{textcomp}
\usepackage{caption}
\usepackage{lscape}

\journal{Preprint}

\begin{document}

\begin{frontmatter}



\title{GANs for Medical Image Synthesis: An Empirical Study}


\author[inst1,inst2]{Youssef Skandarani}
\author[inst3]{Pierre-Marc Jodoin}
\author[inst1,inst4]{Alain Lalande}

\affiliation[inst1]{organization={University of Bourgogne Franche-Comte},
            addressline={}, 
            city={Dijon},
            country={France}}

\affiliation[inst2]{organization={CASIS inc.},
            city={Quetigny},
            country={France}}

\affiliation[inst3]{organization={Department of Computer Science}, 
            addressline={University of Sherbrooke}, 
            city={Sherbrooke},
            country={Canada}}

\affiliation[inst4]{organization={University Hospital of Dijon},
            city={Dijon},
            country={France}}

\begin{abstract}
Generative Adversarial Networks (GANs) have become increasingly powerful, generating mind-blowing photorealistic images that mimic the content of datasets they were trained to replicate.  One recurrent theme in medical imaging is whether GANs can also be effective at generating workable medical data as they are for generating realistic RGB images.  In this paper, we perform a multi-GAN and multi-application study to gauge the benefits of GANs in medical imaging.  We tested various GAN architectures from basic DCGAN to more sophisticated style-based GANs on three medical imaging modalities and organs namely : cardiac cine-MRI, liver CT and RGB retina images. GANs were trained on well-known and widely utilized datasets from which their FID score were computed to measure the visual acuity of their generated images. We further tested their usefulness by measuring the segmentation accuracy of a U-Net trained on these generated images and the original data.  
 Results reveal that GANs are far from being equal as some are ill-suited for medical imaging applications while others are much better off.  The top-performing GANs are capable of generating realistic-looking medical images by FID standards that can fool trained experts in a visual Turing test and comply to some metrics.  However, segmentation results suggests that no GAN is capable of reproducing the full richness of medical datasets.
\end{abstract}




\end{frontmatter}


\section{Introduction}
\label{sec:introduction}
During the last decade, machine learning have been widely adopted mainly due to the advent of deep neural networks and their state-of-the-art  results on a variety of medical imaging tasks. Meanwhile, the introduction of Generative Adversarial Networks (GANs) by \cite{Goodfellow2014GenerativeAN} drove generative modeling and data synthesis to levels of quality never achieved before.  The research on GANs grew at an ever increasing pace, with each iteration pushing back the limits of image quality.  Perhaps, one notable breakthrough in image quality came from \cite{Brock2019LargeSG} and their {\em Big GAN}.  Not so long after,  another dramatic jump in the quality and diversity of generated images came with {\em Style GAN} \citep{Karras2019ASG} which exhibited highly realistic high-resolution human faces.
Motivated by the impressive results achieved by GANs on natural images, the goal of this work is to evaluate how well these machines perform on medical data, an area well-known for its smaller datasets and strict anatomical requirements.

\subsection{Medical image analysis}
Medical image analysis aims to extract uninvasively information about a patient's medical condition. Medical images are images acquired from one of multiple modalities, be it Magnetic Resonance Imaging (MRI), Computed Tomography (CT), Positron Emission Tomography (PET), or Ultrasound (US) to name a few. The acquired images are generally processed using image analysis and/or computer vision techniques to extract certain useful information about the data at hand, for example, to classify whether the case is normal or pathological. One of the most routine tasks in clinical practice is image contouring, or segmentation. Image segmentation is the operation of outlining parts of the images that belong to certain classes of interest.  For example, in the case of cardiac MRI, one may delineate the  left ventricular cavity and myocardium with the objective of measuring blood volumes and contraction rates.

In recent years, machine learning and deep learning garnered a large interest from the medical imaging community due to their unprecedented achievements in a large swath of computer vision tasks.  However, machine learning software have not yet been widely adopted in clinical practice largely due to the fact that neural nets are still error-prone under certain conditions (domain adaptation, different acquisition protocols, missing data, etc).  One reason for this derives from the fact that fully-annotated medical imaging datasets are much smaller than those from other imaging areas.  For example, the gold standard computer vision ImageNet \citep{deng2009imagenet} dataset contains more than 14 million annotated images, while a typical medical image dataset is three to four orders of magnitude smaller. This is because the creation of medical imaging datasets is costly and difficult due to the sensitive nature of the data and the highly specific domain knowledge required to reliably annotate it.
The paucity of training data in medical imaging made the search for other venues of acquiring training sets an active area of research \citep{Frangi2018SimulationAS}.

\begin{figure*}[ht]
\centering
\includegraphics[width=1\linewidth]{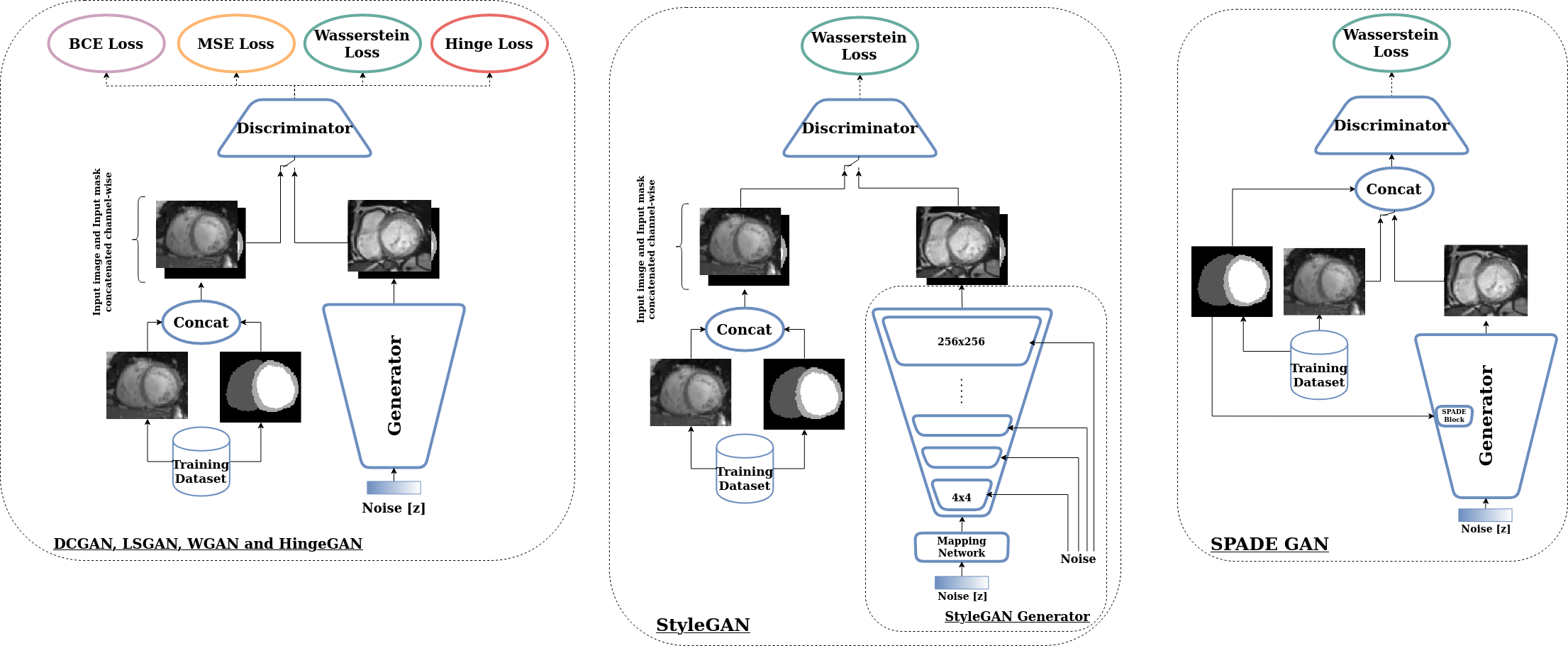}
\caption{\small Architectures of the various GANs we used.}
\label{fig_gans_architecture}
\end{figure*}

\subsection{Synthetic Data and Medical Imaging}
Recently, GANs got a growing attention by the medical research community which used it to synthesis realistic-looking medical images. For example, \cite{Bermudez2018LearningIB} trained a GAN to synthesis new T1-weighted brain MRI with comparable quality as real images and \cite{Baur2018GeneratingHR} succeeded in generating high resolution skin lesion images which experts could not reliably tell apart from real images. \cite{calimeri2017biomedical} took advantage of GANs to generate brain MRI that achieves high scores both in qualitative and quantitative evaluation. In \cite{Chuquicusma2018HowTF}, authors have shown that GAN-generated images of lung cancer nodules are nearly indistinguishable from real images, even by trained radiologists. 

GANs were also used as a mean for generating more training data. In \cite{Shin2018MedicalIS} the authors trained a GAN to generate synthetic brain tumor MRI and evaluated the performance of subsequent segmentation networks trained with the generated data. Looking at the reported results, the segmentation networks trained solely with synthetic data do not come close to those trained with real data performance wise. Likewise, \cite{Skandarani2020OnTE} proposed a combination of a variational autoencoder and a GAN as a data augmentation framework for an image segmentation problem.  Here again, the use of GANs to train downstream neural networks produced mixed (and yet more or less convincing) results.

In addition and as reported in  \cite{Kazeminia2020GANsFM} survey paper, application of GANs in medical imaging extend beyond image synthesis to other tasks, such as domain adaptation, classification, and reconstruction to name a few.
For these applications, the capability of GANs to generate realistic looking images lead to a partial disregard of the usefulness of the generated medical images or whether they hold any value compared to real data in routine clinical tasks.

In the light of these publications, one might wonder how useful GANs truly are in medical imaging.
In this paper, we set out to evaluate the richness and the benefit of using GAN-generated data in the context of medical imaging.  We assess their performances on three datasets of different organs and different modalities.

\section{Generative Adversarial Networks}
\label{sec:gans}
Adversarial networks in general, and GANs more specifically, are trained to play a minimax game between a generator network which tries to maximize a certain objective function in tandem with a discriminator network which tries to minimize that same objective function hence the 'Adversarial' denomination. In their most basic formulation, GANs are trained to optimize the following value function \citep{Goodfellow2014GenerativeAN}:
\begin{equation}
\begin{aligned}
\min_G \max_D V(D, G) = \mathbb{E}_{\bm{x} \sim p_{\text{data}}(\bm{x})}[\log D(\bm{x})] \\ + \mathbb{E}_{\bm{z} \sim p_{\bm{z}}(\bm{z})}[\log (1 - D(G(\bm{z})))].
\end{aligned}
\label{eq:minimaxgame-definition}
\end{equation}

Here, $G(z)$ is the {\em generator network} with parameters $\theta_G$. It is fed with a random variable $z \sim p_{z}$ sampled from a given prior distribution that $G$ tries to map to $x \sim p_{\text{data}}$. To achieve this, another network D (aka the {\em discriminator}) with parameters $\theta_D$ is trained to differentiate between real samples $x \sim p_{\text{data}}$ from a given dataset and fake samples $\hat{x} \sim p_{\theta_G}(x|z)$ produced by the generator. In doing so, the generator is pushed to gradually produce more and more realistic samples with the goal of making the discriminator misclassify them as real.

\subsection{GAN Selection}
The number of papers published on GANs have been growing steadily in the last years.  This has been underlined by \cite{Gonog2019ARG} survey paper which reports no less than 460 references.  Given this large palette of models, we based our choice on those that are the most widely adopted and/or ushered an improvement to the quality of generated images.  We also selected GANs based on their ability to fit on a single 12Gb GPU.

Training GANs can be tricky. Since learning involves two opposing networks, GANs are known for suffering from several training problems, the following three being among the most widely documented.

\textbf{Convergence.}  GANs (and adversarial training in general) often suffer from a lack of defined convergence state.  This is because the training process involves two networks pushing in opposite direction without one out matching the other. This frequently proves to be a difficult task. For example, the generator could become too powerful and learn to fool the discriminator with faulty output.  It could also happen that the discriminator reaches a 50\% accuracy effectively outputting random guesses which does not help the generator learn any meaningful information about the true data distribution.

\textbf{Vanishing Gradients.}  As GANs train a generator with the output of a discriminator, whenever the discriminator significantly outperforms the generator, its loss goes to zero pushing the retropropagated gradient to smaller and smaller value, hence the {\em vanishing gradient} name.  Because of that, the generator does not get enough gradient updates and sees its learning stall to some sub-optimal solutions \citep{Arjovsky2017TowardsPM}.

\textbf{Mode Collapse.}  From all the challenges that obstruct the training of powerful GANs, mode collapse might probably be the most difficult one to deal with.  Mode collapse occurs when the generator gets stuck outputting only one (or a few) modes of the input data distribution. An example could be a generator producing images of  healthy subjects while ignoring the diseased ones. This pitfall leads to a loss of diversity in the generated datasets that can greatly hurt the performance of subsequent networks trained with these generated data.

In regard of the aforementioned criteria and the different challenges, we selected the following GANs for our study.
\subsubsection{DCGAN} Deep Convolutional GANs \citep{Radford2016UnsupervisedRL} are the first GANs to use convolutional layers compared to the inital GAN which used only fully connected layers.  With its simplicity, DCGAN is often the {\em de facto} baseline GAN one implements.  DCGANs showed a considerable jump in image quality and training stability while providing some useful insights on the network design (use of strided convolutions instead of pooling layers, extensive use of batchNorm, etc.).  To our knowledge, DCGAN is among the most widely implemented GAN as of today.
\subsubsection{LSGAN} Least Squares GANs \citep{Mao2017LeastSG} us a different loss for the discriminator than the original GANs which helps alleviate certain challenges and improve the generated sample quality. LSGANs replace the cross entropy loss of the original GAN with the mean squared error which mitigates the vanishing gradient problem leading to a more stable learning process.
\subsubsection{WGAN and WGAN-GP} Wasserstein GANs \citep{pmlr-v70-arjovsky17a} are considered as a major breakthrough to overcome GAN training challenges.  In particular, it is known to reduce the effect of mode collapse and stabilize the learning procedure.  The idea is to use a  Wasserstein earth-mover distance as GAN loss function together with some other optimization tricks like weight clipping and gradient penalty.
\subsubsection{HingeGAN (Geometric GAN)} Introduced by \cite{Lim2017GeometricG}, HingeGANs substitute the original GAN loss for a margin maximization loss which theoretically converges to a Nash equilibrium between the generator and discriminator.  As for WGAN and LSGAN, HingeGAN has the sole benefit of easing the optimization process.
\subsubsection{SPADE GAN} Spatially Adaptative DEnormalization (SPADE) GANs \citep{Park2019SemanticIS} is a member of the so-called {\em image-to-image} translation GAN family.  SPADE GANs produce state-of-the-art  results on a wide range of datasets producing high quality images perfectly aligned to a semantic input mask. SPADE GANs come as an improvement of the previously-published pix2pix \citep{Isola2017ImagetoImageTW} model. 
\subsubsection{Style based GANs} StyleGAN \citep{Karras2020AnalyzingAI}, often considered as the state-of-the-art generative neural network, introduces multiple tricks to GAN borrowed from previous works such as progressive GANs \citep{karras2018progressive} that gradually trains the GAN with different resolutions which leads to better quality and a more stable training process. StyleGAN also comes with a greatly modified generator which includes adaptive instance normalization blocks (AdaIN), the injection of noise at every level of the network and use a 8-layer MLP mapping function on the input latent vector $\vec z$.

\subsection{Evaluation Metrics}
Broadly speaking, the metrics used to quantify the effectiveness of GANs are the same as those used to evaluate traditional image synthesis tasks. This boils down to computing a similarity distance between a set of images. In their early stages, GANs were evaluated using the traditional metrics such as {\em Peak Signal to Noise Ratio} (PSNR) \citep{Regmi2018CrossViewIS} or {\em Structural Similarity Index Measure} (SSIM) \citep{Odena2017ConditionalIS}. As the field advanced, more image quality metrics emerged and became the {\em de facto} evaluation criteria, such as {\em Learned Perceptual Image Patch Similarity} (LPIPS) \citep{Zhang2018TheUE}, {\em Inception Score} (IS) \citep{Salimans2016ImprovedTF} and the {\em Frechet Inception Distance} (FID) \citep{Heusel2017GANsTB}.

First introduced by \cite{Heusel2017GANsTB}, the {\em Frechet Inception Distance} (FID), makes use of a pretrained Inception network on the ImageNet \citep{deng2009imagenet} dataset to assess the quality of GAN generated images. The FID is a distance between the distribution of the GAN sampled images and the real dataset used to train the GAN. Generated samples and real images are fed to the pretrained Inception network and the mean and covariance of the activations in the final block, assumed to be of a Gaussian distribution,  are collected for both sets then the Frechet distance is computed between both. The FID is computed on a learned feature space and was shown to correlate well to human visual perception~\citep{Zhang2018TheUE}. Although, it still suffers from a number of drawbacks \citep{Borji2019ProsAC}, most prominently, it suffers from a high bias \citep{Binkowski2018DemystifyingMG}. Also, FID can not detect a GAN that memorizes the training set \citep{Lucic2018AreGC}.

In this work, we use the FID metric as it evolves in tandem with human perception. In addition, it makes use of the original dataset to compute a distance in a learned feature space.

\begin{figure}[tp]
\centering
\includegraphics[width=1\linewidth]{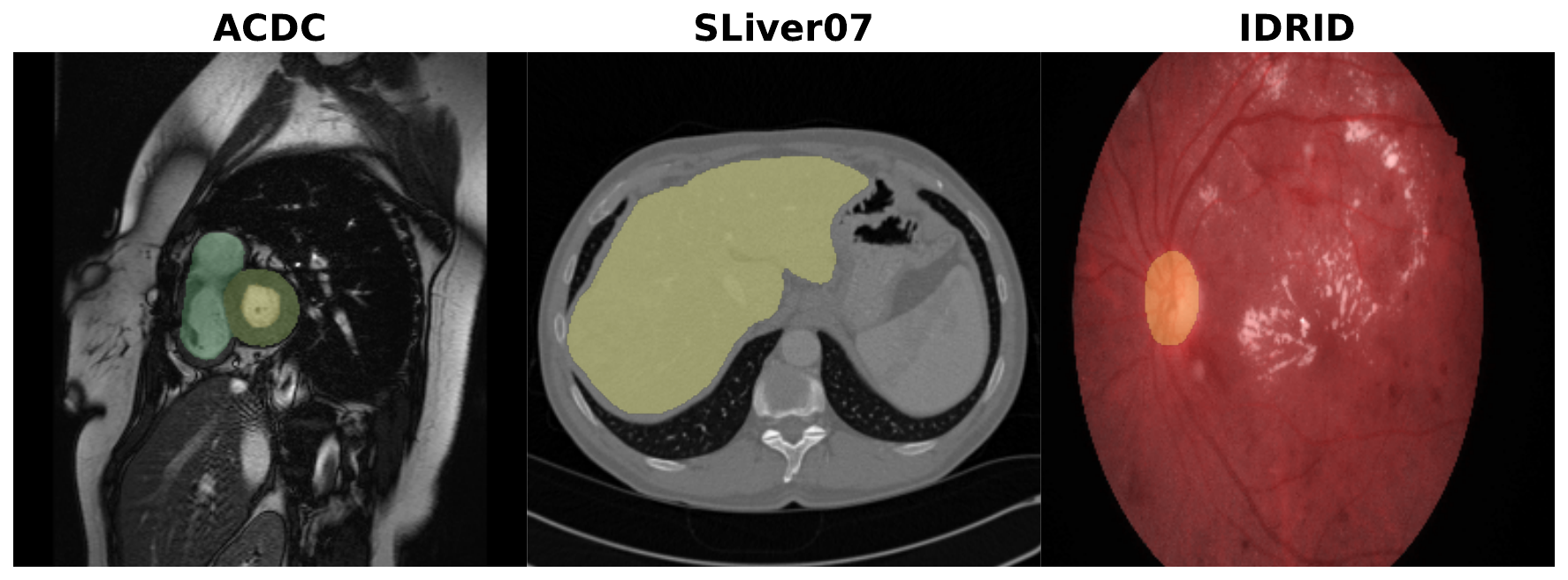}
\caption{\small Examples of images and the segmented structures for ACDC, SLiver07 and IDRID datasets.}
\label{fig_orig_images_ex}
\end{figure}

\section{Material and Methods}
\label{sec:protocol}
To make informed decisions about the usefulness of GANs in medical imaging as a source of synthetic data, we had to take into account different GANs and cover a diverse set of image modalities. In parallel, a wide range of hyperparameters had to be covered to assess their effect on the GANs at hand.

\subsection{Hyperparameters search}

GANs are known for their sensitivity to hyperparameters tweaking \citep{Lucic2018AreGC}. In order to achieve a fair comparison between the selected GANs, we covered a wide spectrum of hyperparameters (some affecting the GAN architecture), through a vast hyperparameter search, totaling roughly 500 GPU-days. We retained the best performing runs with regards to the reference metric FID for its correlation with subjective evaluation.

Moreover, since the number of runs needed to sweep a large hyperparameter space grows exponentially with the number of hyperparameters we set to optimize over, we choose a number of sensible initial configurations for each dataset/GAN pair mostly based on their default configuration. 
Table \ref{tab_hparams} lists the hyperparameters we searched over. Iterating over these hyperparameters enabled us to find the set that works best for each GAN/dataset pair. In addition, this hyperparameters search also gave us a peek at how the training stability is affected by the selected hyperparameters. Note that some combinations were only tested for specific GANs, such as "weight clipping" for WGAN or "Gradient Penalty" for WGAN-GP.

\subsection{GANs setup}
The training of the DCGAN, LSGAN, WGAN and HingeGAN followed the same protocol. A traditional fully convolutional network architecture with a standard generator and discriminator composed of upconvolutions and strided convolutions respectively was implemented, as a basis of our DCGAN. Then the loss function was swapped to convert it to either a LSGAN, a WGAN or a hingeGAN. For StyleGAN and SPADE GAN, we relied on the publicly available implementations without any change to the networks architecture.
Fig. \ref{fig_gans_architecture} schematically summarize  the architecture of each GAN.

\begin{table*}
\small
\centering
\resizebox{\textwidth}{!}{\begin{tabular}{l|c}
\hline
\textbf{Hyperparamters}             &   Values                          \\
\hline
Differentiable Augmentation \citep{Zhao2020DifferentiableAF}    & TRUE/FALSE        \\
Activation fn of Discriminator         & ReLU/LeakyRelu/Elu/Selu                     \\
Activation fn of Generator             & ReLU/LeakyRelu/Elu/Selu    \\
Normalization layer of Discriminator   & BatchNorm \citep{Ioffe2015BatchNA}/  InstanceNorm \citep{Ulyanov2016InstanceNT} \\
Normalization layer of Generator       & BatchNorm \citep{Ioffe2015BatchNA}/  InstanceNorm \citep{Ulyanov2016InstanceNT} \\
Number of Filters of Discriminator     & 16/32/64/128  \\
Number of Filters of Generator         & 16/32/64/128  \\
Use Spectral norm for Discriminator    &  TRUE/FALSE   \\
Use Spectral norm for Generator        &  TRUE/FALSE   \\
Weight initialization function          & Normal/Xavier/Xavier Uniform/ Kaiming He \\
Weight initialization Gain              & 0.01/0.02/0.1/1.0 \\
Gradient Penalty loss weight (WGAN-GP only) & 0/0.1/1.0/10.0 \\
Weight Clipping Value (WGAN only)       & 0 / 0.01 / 0.1 \\
Feature matching loss weight           & 0 / 1.0 /10.0 \\
VGG loss weight                        & 0 / 1.0 /10.0 \\
Learning Rate                          & 0.00004 / 0.00005 / 0.0001 / 0.0002 / 0.001 \\
Use of Label Smoothing \citep{Salimans2016ImprovedTF}                & TRUE/FALSE      \\
Use of Data Augmentation                &  TRUE/FALSE           \\
\hline
\end{tabular}}
\caption{List of the different hyperparameters we optimized over.}
\label{tab_hparams}
\end{table*}

\subsection{GAN Training Tricks}
In order to make the GAN training process more stable, we rely on a few tricks that have been shown useful in this regard.

\textbf{Label smoothing.} First applied to GANs by \cite{Salimans2016ImprovedTF}, it consists in replacing the true classification labels given to the discriminator to a smooth value $\alpha$.

\textbf{Feature matching.} Also introduced by \cite{Salimans2016ImprovedTF}, feature matching adds another objective to the generator of the GAN which consists in minimizing a distance between the activations of the discriminator for real and generated data.

\textbf{Differentiable augmentation.} Presented by \cite{Zhao2020DifferentiableAF}, differentiable augmentation imposes various types of augmentation on the fake and real samples fed into the discriminator yielding a more stable training and a better convergence.

\subsection{GAN evaluation in medical imaging}
While image fidelity is fundamentally important for a practitioner to deliver a good diagnostic, the visual acuity of generated images cannot be the sole marker to assess the true performances of GANs.  In this paper, we want to assess how rich and diverse a synthetically generated dataset really is in the context of medical imaging.  

Thus, to verify the medical viability of GAN-generated images, we trained independently a second network as a downstream task on the GANs-generated datasets and compared its results to those obtained on the original (yet real) datasets.  In this work, the downstream task is segmentation.

This assessment sets a common evaluation protocol for every GAN.  This evaluation is also insightful considering that the objective for using GANs is often to artificially increase the size of a dataset and thus provide more training data to a subsequent task~\citep{Skandarani2020OnTE, Shin2018MedicalIS}. This approach has been explored before with GANs trained on natural images and evaluated through a classification task~\citep{Ravuri2019ClassificationAS,Shmelkov2018HowGI}.

\subsection{Datasets}

To cover a good spectrum of image and medical applications, we picked up three different datasets based on their imaging modalities, their organ of interest and size, namely cardiac cine-MR images, liver CT and retina imaging. These datasets offer a varied selection of data. Different dataset sizes are present from large (Sliver07) to moderate (ACDC) to small (IDRID). Coupled with that, different image modalities and organ shapes are considered. Fig. \ref{fig_orig_images_ex} shows an example of images for each of the datasets.
\subsubsection{ACDC}
The {\em Automated Cardiac Diagnosis Challenge} (ACDC) dataset~\citep{bernard2018deep} consists of 150 exams (100 training and 50 testing) of short-axis cardiac cine-MRI acquired at the University Hospital of Dijon (all from different patients).  The exams are divided into 5 evenly distributed subgroups (4 pathological plus 1 healthy subject groups) and further split into 100 exams (1902 2D slices) for training and 50 exams (1078 2D slices) held out set for testing. The pixel spacing varied from $0.7 mm$ to $1.9 mm$ with a slice spacing between $5 mm$ to $10 mm$. The exams come with multi structure segmentation masks for the right ventricular cavity, the left ventricular cavity and the left ventricular myocardium at end-diastole and end-systole times.
\subsubsection{SLiver07}
The {\em Segmentation of the Liver Competition} 2007 (SLIVER07) \citep{StynerSliver07} dataset contains 40 CT volumes of the liver enhanced with contrast agent.  Most livers are pathological and include at least one tumor. The pixel spacing ranges from $0.55 mm$ to $0.8 mm$ and the inter-slice gap between $1 mm$ to $3 mm$.
The 40 CT datasets are randomly split in three groups: a group of 20 volumes for training, another group of 10 volumes for validation and remaining 10 volumes for testing. For our study, we only use on the 20 training volumes provided with manual segmentations for the liver which totals 4159 2D slices.
\subsubsection{IDRID}
The {\em Indian Diabetic Retinopathy Image Dataset} (IDRiD) \citep{Porwal2018IndianDR} contains a total of 516 retinal fundus images of normal and pathological cases. Images are provided with disease grading ground truth for the full dataset and segmentation masks for 81 images. We used part of the 81 images for our study, specifically the 54 training images with the optical disc segmentation masks.

\subsection{Dataset generation}

For our study, our selected GANs were trained on the aforementioned datasets with the goal of synthesizing new medical data.  The overarching objective of this study is to assess whether or not GANs offer a reliable framework for synthesizing realistic and diverse medical images. To examine how well GANs manage to learn the original data distribution, a large number of images was sampled from each of our trained GANs which we later used to train a segmentation network.

To be able to train a downstream segmentation network, the different GANs were trained on the joint distribution of the image and the mask by concatenating the channel axis.  We did so for every GAN except for SPADE which is by nature conditioned on a segmentation mask.
Once properly trained with the right set of hyperparameters, each GAN was used to generate a dataset of 10,000 images by randomly sampling the input latent space. No further processing was done on the generated datasets as the objective was to gauge the quality of the {\em raw} images output by the GANs. Fig. \ref{fig_gen_images_ex} shows some examples of images generated by each GAN on the three datasets.

\begin{figure}[tp]
\centering
\includegraphics[width=0.8\linewidth]{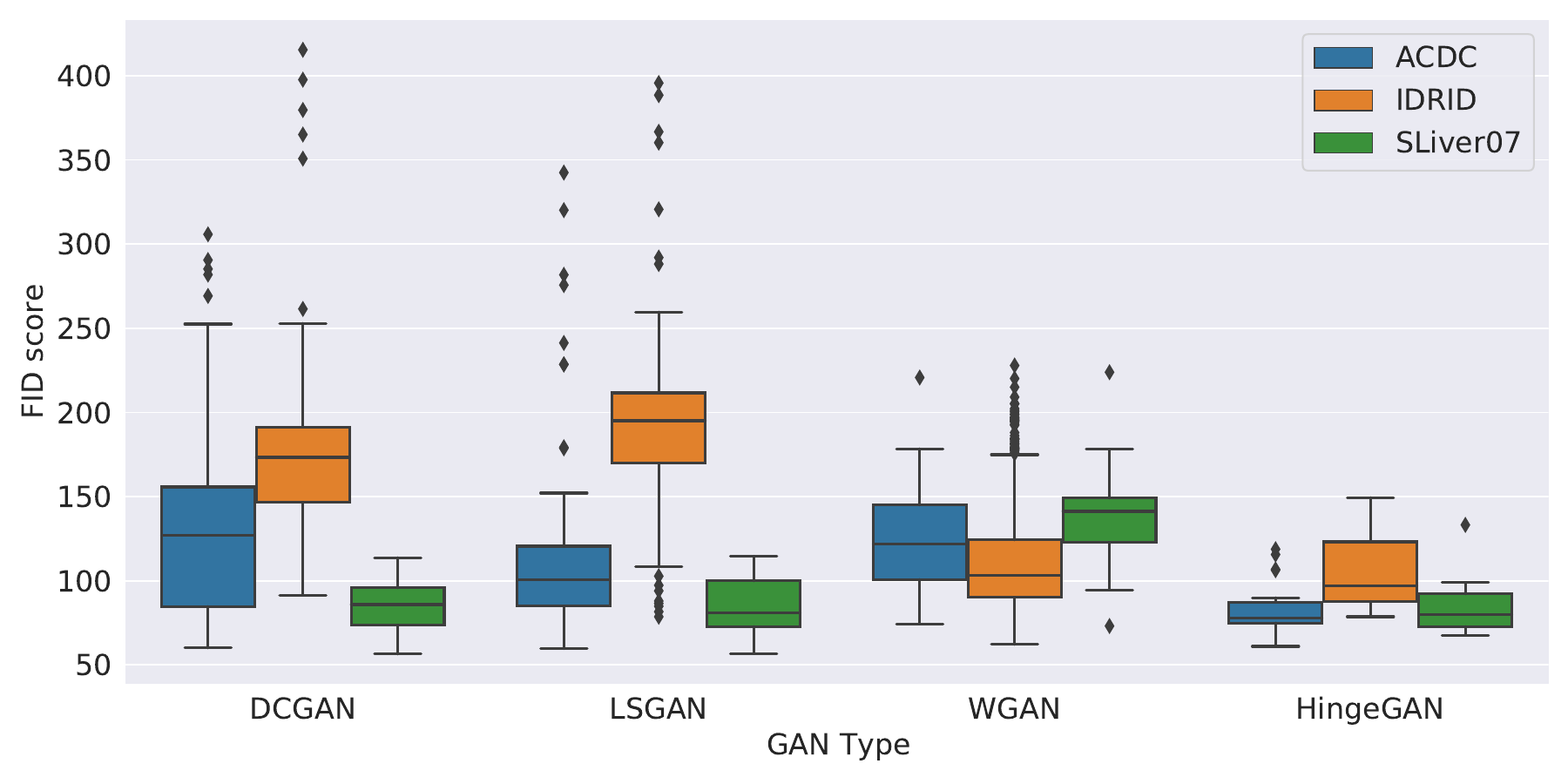}
\caption{\small FID Score for different GAN types on IDRID, ACDC and SLiver07 Datasets across different hyperparameters settings.}
\label{fig_fid_loss_acdc}
\end{figure}

\section{Experiments and Results}
\label{sec:experiments}
This section goes through the experiments and results obtained by each GAN on each dataset.  

\subsection{Hyperparameter search and overall results}

The hyperparameter search performed on DCGAN, LSGAN, WGAN and HingeGAN revealed interesting insights.  The first one is that some GANs are very sensitive to their hyperparameters.  To underline this, the FID score obtained for every set of hyperparameters of each GAN and each dataset are shown on the Fig. \ref{fig_fid_loss_acdc}.  As one can see, the HingeGAN has the lowest variance and, overall, the best FID score.  On the other hand, DCGAN and LSGAN are overall much more sensitive to hyperparameter tweaking.  This is inline with our qualitative experience as the training of DCGAN and LSGAN often ended up producing degenerated images.   As for SPADE and Style GAN, they were not included in the graph due to the shear amount of training time they require (it took respectively 10 days and 30 days to train them) but also due to their remarkable stability.  Empirical evidence obtained with different hyper-parameters on a few epochs  suggest that their FID variance is much lower than that of HingeGAN hence why they ended up with top results with almost no hyperparameter tweaking.

\begin{figure}[tp]
\centering
\includegraphics[width=0.7\linewidth]{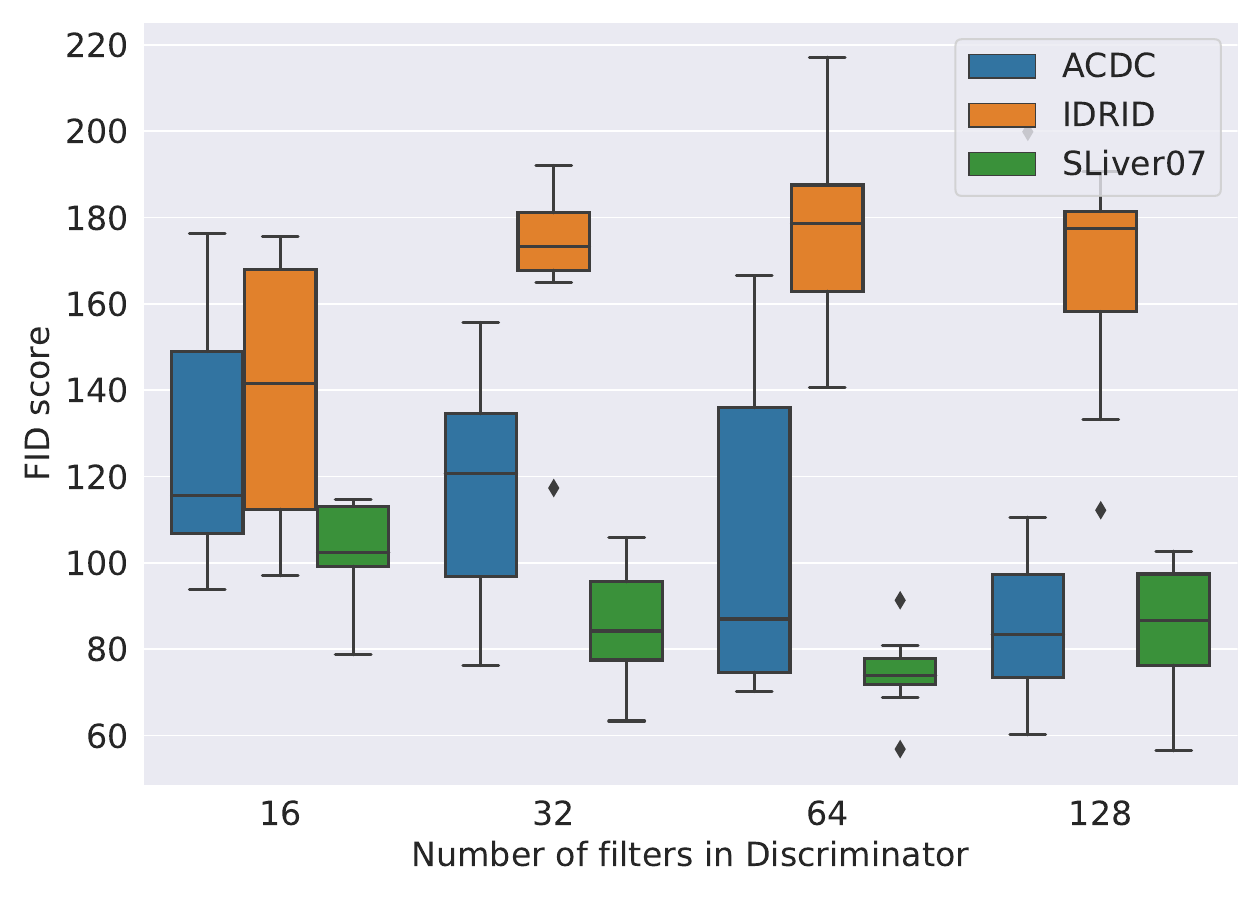}
\caption{\small FID Score for different number of filters for the discriminator of the DCGAN, LSGAN, WGAN and HingeGAN.}
\label{fid_hparams}
\end{figure}

Another insight comes from the impact a dataset has on the performances of GANs.  As can be seen from Fig. \ref{fig_fid_loss_acdc}, the larger the reference dataset is, the better the resulting FID will be.  It goes from IDRID, the smaller one with FID values well above 150.  Then ACDC with FIDs values roughly between 100 and 150 and finally Sliver07, the largest dataset in number of images, with most FID values below 100.  A similar trend can be seen in Fig.~\ref{fid_hparams} where the overall FID values for every GANs is shown against the number of convolutional filters in the discriminator network.  This shows how volatile GANs can be when trained on smaller datasets such as IDRID.  Similar plots with other hyperparameters can be found in the supplementary materials.

The best FID score obtained for each GAN and each dataset is shown in  the third column of Table~\ref{tab_fid_dice}.  Example of generated images can also be seen in Fig.~\ref{fig_gen_images_ex} (and in high resolution in the supplementary material).  By far, the two best models are StyleGAN and SPADE GAN.   The most extreme case is for IDRID where a SPADE GAN got a surprising FID of $1.09$ and remarkably vivid images in  Fig.~\ref{fig_gen_images_ex}.

\begin{table}
\centering
\caption{\small FID and U-Net Dice score for different GANs on ACDC, IDRID and Sliver07 datasets.}
\begin{tabular}{l|c|c|c} 
\hline
Dataset & GAN       & FID score & U-Net Dice score \\ 
\hline
& Original Data & --   & \textbf{0.89}  \\ 
& Augmented Original Data & --   & \textbf{0.90}  \\
\hline
& DCGAN     & 60.12     & 0.30  \\ 
& LSGAN     & 59.65     & 0.39  \\ 
ACDC & WGAN      & 74.30     & 0.70 \\ 
& Hinge GAN & 61.00     & 0.63  \\ 
& SPADE GAN & 41.54     & 0.86  \\ 
& StyleGAN  & \textbf{24.74}     & \textbf{0.87}  \\
& Orig. Data + SPADE GAN & --     & \textbf{0.90} \\ 
& Orig. Data + StyleGAN  & --     & \textbf{0.90}  \\
\hline
\hline
& Original Data & --   & \textbf{0.83}  \\ 
& Augmented Original Data & --   & \textbf{0.84}  \\ 
\hline
& DCGAN     & 91.34     & 0.29  \\ 
& LSGAN     & 78.61     & 0.20  \\ 
IDRID & WGAN      & 62.12     & 0.72 \\ 
& Hinge GAN & 78.61     & 0.69  \\ 
& SPADE GAN & \textbf{1.09}     & \textbf{0.82}  \\ 
& StyleGAN  & 23.72     & 0.80   \\
& Orig. Data + SPADE GAN & --     & \textbf{0.84}  \\ 
& Orig. Data + StyleGAN  & --    & \textbf{0.84}  \\
\hline
\hline
& Original Data & --   & \textbf{0.72}  \\
& Augmented Original Data & --   & \textbf{0.70}  \\
\hline
& DCGAN     & 56.41     & 0.14  \\ 
& LSGAN     & 56.82     & 0.15  \\ 
Sliver07 & WGAN      & 73.11     & 0.16 \\ 
& Hinge GAN & 67.69     & 0.15  \\ 
& SPADE GAN & 47.62     & \textbf{0.61}  \\ 
& StyleGAN  & \textbf{29.06}     & 0.36  \\
& Orig. Data + SPADE GAN & --     & \textbf{0.71}  \\ 
& Orig. Data + StyleGAN  & --     & \textbf{0.71}  \\
\hline
\end{tabular}
\label{tab_fid_dice}
\end{table}

\begin{figure*}[ht]
\centering
\includegraphics[width=1\linewidth]{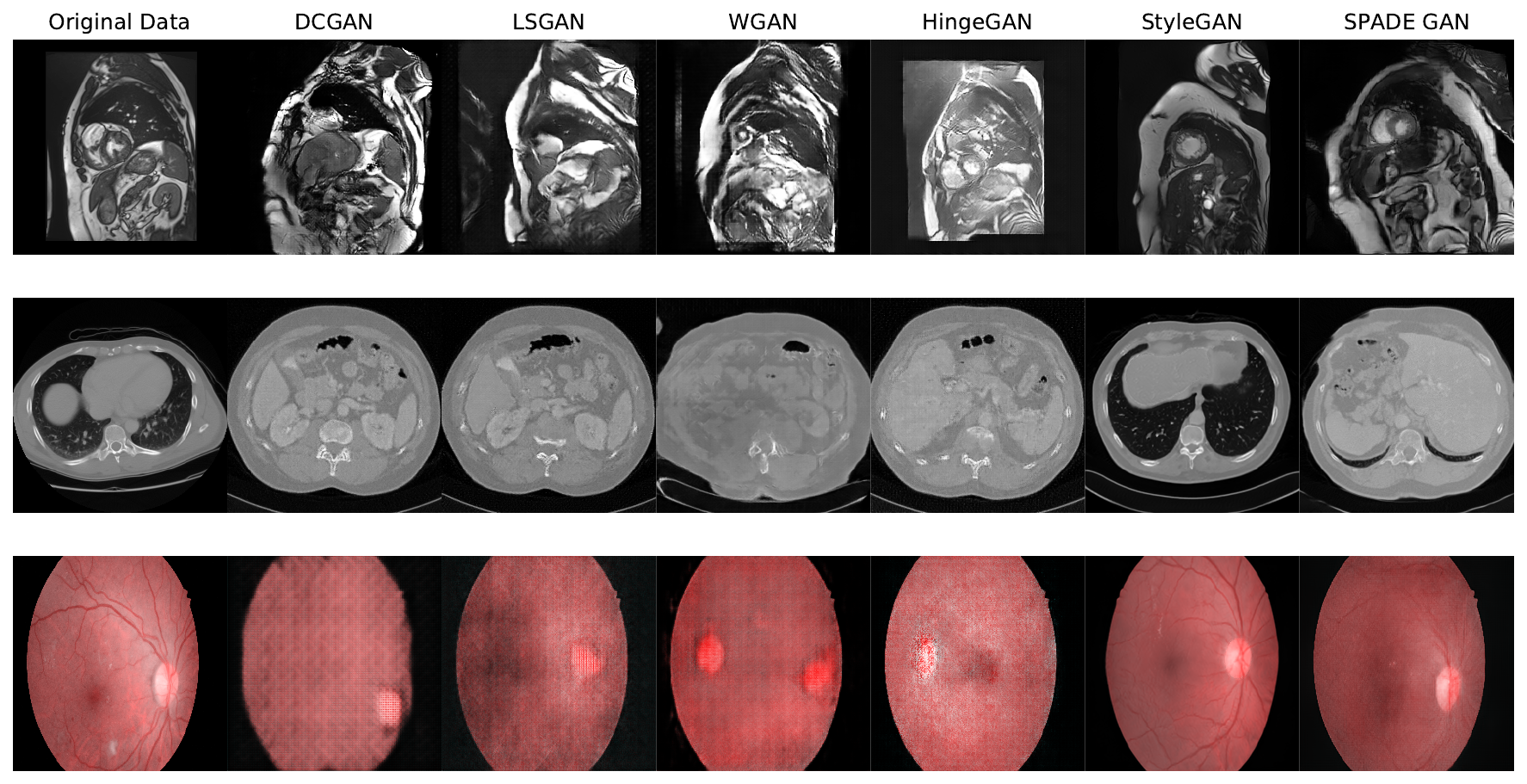}
\caption{\small Examples of generated images for each GAN on the ACDC, SLiver07 and IDRID datasets. The first column is an example of image from the real dataset.  High resolution versions of these images are available in the supplementary material.}
\label{fig_gen_images_ex}
\end{figure*}
\subsection{Segmentation evaluation}
\label{sec:seg_eval}

\begin{figure}[h]
\includegraphics[width=1\linewidth]{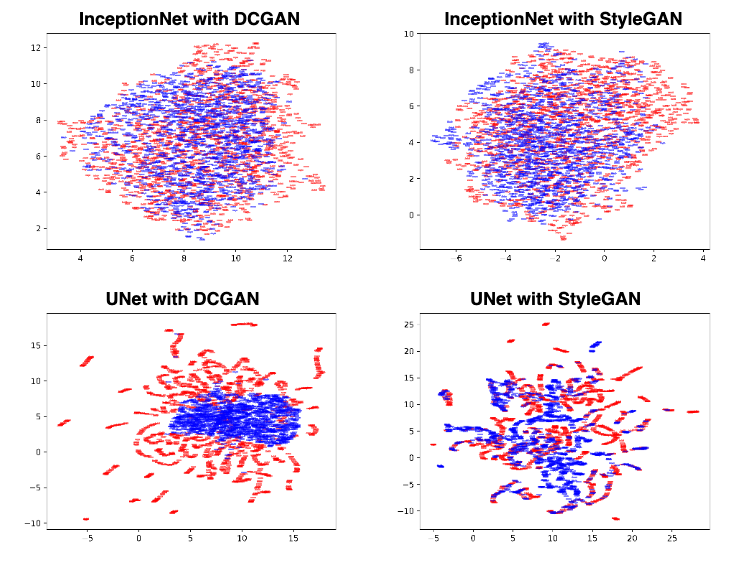}
  \caption{\small Comparison of UMap projection of activations of images generated by a DCGAN and others generated by a StyleGAN, with an InceptionNet trained on ImageNet (top row) and a U-Net trained on the original dataset (bottom row). Red points: Real images, Blue points: Generated images }
\label{fig_umap_comparison}
\end{figure}

As mentioned before, the true value of the generated images was validated with a downstream segmentation network trained on the synthetic data instead of the original (yet real) data.  To do so, $10,000$ new images were generated for each dataset and a U-Net~\citep{ronneberger2015u} was trained to predict the segmentation mask.  

The last column of Table \ref{tab_fid_dice} contains the Dice score obtained on the {\em real} test set of each dataset.  Unsurprisingly, as suggested by the FID scores, StyleGAN and SPADE GAN achieve the highest Dice scores on all the datasets with StyleGAN reaching 87\% Dice on the ACDC dataset, 2\% less than when training with the original data. 

These results reveal three important things about GANs in medical imaging.  
First, simpler models such as DCGAN, LSGAN, WGAN, and HingeGAN perform systematically poorly on every dataset despite intensive hyperparameter search.  This suggest that these models might be ill-suited for medical imaging applications. 

Second, despite their visual similarity, GAN-generated datasets do not have the same richness as real datasets.  This is illustrated by the fact that despite being trained on far more images, none of the GAN Dice score equals or outperforms the ones obtained on the original datasets. Moreover, the generated datasets, when used as augmentation data, achieve similar performance to traditional augmentation techniques (rotations, shifts, flips) illustrated by the Dice score of training with a mix of the original data and generated data and the augmented original data only.

Third, while the FID score is a good proxy to distinguish the {\em best} methods from the {\em least effective} ones, it does not correlate well with an application score such as Dice.  For example, the FID score of 29.06 of StyleGAN on Sliver07 suggest that the produced images are much more accurate than those of SPADE GAN (FID=47.62).  However, the resulting Dice scores show that SPADE GAN is significantly better than any other model.  A similar comment can be made for IDRID and ACDC as StyleGAN and SPADE GAN got similar Dice score but very different FIDs.
 As for the FID score of 1.09 obtained by SPADE GAN, the associated 82\%  Dice score suggest that the network has most likely memorized the training set. This might be attributed to the small size of the IDRID dataset as well as the simple shape of the input segmentation mask.

To further analyze whether the FID score is a reliable medical imaging metric, we plotted the InceptionNet latent space of the  generated images obtained with the most and the least effective GANs, i.e. DCGAN and StyleGAN (c.f. top row of Fig.~\ref{fig_umap_comparison}, plots were obtained with UMap \citep{McInnes2018UMAPUM}).  
In parallel, we plotted the U-Net latent space for the same images and the same GANs (cf. bottom row of Fig. \ref{fig_umap_comparison}). While the red and the blue InceptionNet scatter plot distributions are very similar for DCGAN and StyleGAN the U-Net ones reveal much more distinctive patterns.  Indeed, the U-Net distributions of  StyleGAN follow very similar distributions (hence suggesting that the syntehtic images of StyleGANs are visually very close to those of the original dataset) while the ones from DCGAN show a clear case of mode collapse.  This underlines a fundamental limit of the FID metric: since the InceptionNet was trained on ImageNet (a non-medical dataset) its use in medical imaging must be made with great care.

\subsection{Visual Turing Test}
Considering how realistic looking some of GAN-generated images are, we asked four medical experts, each with more than 15 years of experience in cardiology, to classify fake and real cine MRI images generated by StyleGAN and the ACDC dataset.  Each expert was shown $100$ images consisting of a 50/50 mixture of real and synthetic images and were asked to classify it based only on their visual appreciation.  The experts managed to achieves an average accuracy of only 60\% thus showing how visually accurate the generated images are.

\section{Discussion}
\label{sec:discussion}
In this section, we go through the aspects that play a major role in the process of training GANs with medical data.

\subsection{Training Volatility}
Throughout this work, the training instability of GANs was a recurrent theme underlying how slight hyperparameter adjustments can considerably affect the training process. In contrast, GANs were not equally sensitive to the selected hyperparameters. While it is true that DCGAN and LSGAN showed the highest variability, it came to be easier to train WGAN and HingeGAN which were less sensitive to hyperparameters selection.

Moreover, even though the state-of-the-art GANs, such as SPADE or StyleGAN, seem to be the only viable pick to produce images of high quality, they still suffer from long training times and can sometimes lead to overfitting and "Memory GAN", i.e. a GAN that outputs the training set. 

Likewise, in the case of the smaller GANs, finding the right set of hyperparameters was not always simple. To illustrate this point, we went through a total of 1,500 training runs with different hyperparameter combinations. Most of the runs lead to models that could not generate meaningful images, while the remaining runs did not always fair well when evaluated with the FID or through the image segmentation task.
Concurrently, although a considerable amount of hyperparameters were explored, we have not had enough GPUs to go through a GAN architecture search which could have provided better performances.

\subsection{FID and Image Quality}
We relied on the FID score to monitor the training of the GANs. We also compared FID to a domain specific evaluation (segmentation Dice score). This process, enabled us to better understand to what extent a FID metric optimized for natural images can be used in medical imaging. Our results reveal that the FID score continuously improves as the training of any GAN moves forward.  In contrast, the FID score could not be consistently relied on as a measure of the image quality when used as training input for subsequent tasks. Table \ref{tab_fid_dice} clearly shows that lower FID does not always yield better performance on a subsequent task of image segmentation.
These results make it interesting to ask whether metrics grounded in domain specific knowledge could help make GANs easier to evaluate and compare.

\subsection{Data Scale}
When comparing the results on the three datasets, an important trend related to the performance of the GANs and the data is visible. Indeed, when the size of the input dataset is exceedingly small, as is the case for the IDRID dataset in our study, the expected benefit of training a GAN to increase the dataset size quickly dissipates as they often overfit which can have the adverse effect on the subsequent task. 
In parallel, when the input dataset is highly unbalanced, portrayed by the SLiver07 dataset in our study with only 5\% foreground pixels, the trained GANs can further exacerbates this imbalance as they will ultimately learn the underlying biases of the training data. Moreover, most of the prevalent GAN architectures deal with 2D images, which is not necessarily the best format to train with when dealing with medical imaging data as it might have been acquired in 3D. This might explain further the bad performance shown on the liver CT dataset.

\subsection{Compute Scale}
One shall keep in mind that training a GAN is often computationally intensive (typically because it involves two or more networks) and require a large amount of memory.  Also, training GANs requires a lot of hyperparameter tuning which may or may not lead to better results  when considering the downstream tasks the generated data is intended for. This also affects more sophisticated GANs which, despite their good performances which can fool medical experts, require large computing resources to train. For example, the StyleGAN took roughly 30 days to train on the ACDC dataset with a NVidia Titan V GPU with 12Gb of memory.  And yet, StyleGAN did not always offer a guarantee to the usefulness of the generated samples (Dice score of 0.36 for StyleGAN on the Sliver07 dataset).

\subsection{Medical worth}
As there is no automated objective way to assess whether a medical image conveys the information for the diagnosis it is intended for, we based our analysis on a proxy task that aims to mimic the process for which a dataset is created, and compared its performance to that of the original data. The results show that, although most of the images generated by the tested GANs fail in reaching the baseline performance, some of the more advanced ones manage to close the gap. However, when subjectively assessing the images generated by the larger GANs, we can still see that they exhibit a remarkable degree of complexity and quality. This might be related to the smaller scale of the datasets in medical imaging and the difference in their nature with the original datasets for which most of the GANs were tailored. Likewise, a considerable amount of the medical data is acquired in a 3D fashion and voxel wise, e.g. CT. Typical GANs might not capture the full extent of the medical information when trained solely on 2D views.
Indeed, this makes exploring GANs specially made for medical data an interesting research venue and could lead to an improvement in quality and ultimately clinical usability.

\section{Conclusion}
\label{sec:conclusion}
Currently the use of deep learning approaches in medical image analysis stay hindered by the limited access to huge annotated dataset. To address this limitation, we have probed both the limitations and promising aspects of Generative Adversarial Networks as medical image synthesis tools through an experimental approach on three different datasets. As a result, GANs effectiveness as a source of medical imaging data was found to be not always reliable, even if the produced images are nearly indistinguishable from real data. Tangentially, results point that traditional metrics used to evaluate GANs are less robust than task based evaluations.

All in all, this study should drive more research in GANs that take into account the different subtleties of medical data and hopefully lead to better generative models.

\section*{Acknowledgment}
We would like to acknowledge Nathan Painchaud and Carl Lemaire for their help with implementations and computing resources. We would like to thank Kibrom Berihu Girum and Raabid Hussain for useful discussions and remarks.

\section*{Conflict of interest statement}
The authors declare that they have no known competing financial interests or personal relationships that could have appeared to influence the work reported in this paper.


\footnotesize
 \bibliographystyle{elsarticle-harv} 
 \bibliography{main}





\end{document}